\documentclass[12pt]{article}
\usepackage{natbib,amssymb}
\usepackage{graphicx}

\begin{document}

\title{\bf Tidal decay of circumbinary planetary systems}

\author{Ivan I. Shevchenko}

\author{Ivan~I.~Shevchenko\/\thanks{E-mail:~ivan.i.shevchenko@gmail.com} \\
Institute of Applied Astronomy, RAS, \\ 191187 Saint Petersburg, Russia \\
Lebedev Physical Institute, RAS, \\ 119991 Moscow, Russia}

\date{}

\maketitle

\begin{center}
Abstract
\end{center}

\noindent It is shown that circumbinary planetary systems are
subject to universal tidal decay (shrinkage of orbits), caused by
the forced orbital eccentricity inherent to them. Circumbinary
planets (CBP) are liberated from parent systems, when, owing to
the shrinkage, they enter the circumbinary chaotic zone. On
shorter timescales (less than the current age of the Universe),
the effect may explain, at least partially, the observed lack of
CBP of close-enough (with periods $< 5$~days) stellar binaries; on
longer timescales (greater than the age of the Universe but well
within stellar lifetimes), it may provide massive liberation of
chemically evolved CBP. Observational signatures of the effect may
comprise (1)~a prevalence of large rocky planets (super-Earths) in
the whole population of rogue planets (if this mechanism were the
only source of rogue planets); (2)~a mass-dependent paucity of CBP
in systems of low-mass binaries: the lower the stellar mass, the
greater the paucity.

\bigskip


\bigskip

\section{Introduction}

Consider a planar three-body system of gravitating bodies: a
central massive binary and a much less massive particle orbiting
around the binary. Thus, the particle's orbit is circumbinary. The
orbital eccentricities of circumbinary particles are periodically
forced on secular and local orbital timescales
\citep{MN04,PLT12,DS15,AIR17}. Therefore, the circumbinary orbits
cannot be permanently circular. This phenomenon provides a natural
universal mechanism of internal tidal friction and heating in
circumbinary planets (CBP) \citep{S17AJ}. In this article, we
describe and consider the planetary escape process that takes
place as a result of this shrinkage. Indeed, a particle in the
slowly shrinking circumbinary orbit enters eventually the chaotic
zone around the central binary, and therefore escapes. Once in the
zone, the particle escapes inevitably, being subject to the
chaotic diffusion along the ``staircase'' of overlapping integer
mean-motion resonances (between the binary and the particle), up
to crossing the separatrix between the bound and unbound dynamical
states \citep{S15}.

We show that the effect of tidal decay may explain, at least
partially, the observed lack of CBP of close-enough (with periods
$< 5$~days) stellar binaries. What is more, on longer timescales
(greater than the age of the Universe but well within stellar
lifetimes), it may provide massive liberation of chemically
evolved CBP.

\section{Tidal heating and orbital decay}

For brevity, we consider the case of equal-mass stellar binaries
solely. It can be shown that moderate deviations from the mass
equality do not influence the derived timescales much. In the
equal-mass case, the secular term in the time behavior of CBP's
eccentricity vanishes; however, the eccentricity is still forced
on a shorter timescale \citep{MN04,PLT12}. By averaging the
perturbing function on the orbital timescale of the central
binary, and retaining the terms up to the second order in the
planetary eccentricity, \cite{PLT12} found that the eccentricity
of a CBP (put initially in a circular orbit around the binary)
oscillates around the forced non-zero value

\begin{equation}
e_\mathrm{ff} = \frac{3}{4} \frac{m_1 m_2}{(m_1 + m_2)^2}
\left(\frac{a_\mathrm{b}}{a_\mathrm{p}}\right)^2 \left( 1 +
\frac{34}{3}e_\mathrm{b}^2 \right)^{1/2} , \label{eq_e_ffP}
\end{equation}

\noindent where $m_1$ and $m_2$ are the masses of the binary
components, $e_\mathrm{b}$ is the eccentricity of the binary, and
$a_\mathrm{b} \ll a_\mathrm{p}$ are the semimajor axes of the
binary and the CBP, respectively. Equivalently, one has

\begin{equation}
e_\mathrm{ff} = \frac{3}{4} \mu (1 - \mu)
\left(\frac{T_\mathrm{b}}{T_\mathrm{p}}\right)^{4/3} \left( 1 +
\frac{34}{3}e_\mathrm{b}^2 \right)^{1/2} , \label{eq_e_ff}
\end{equation}

\noindent where $\mu = m_2/(m_1 + m_2)$ is the mass parameter
($m_1 \geq m_2$), and $T_\mathrm{b}$ and $T_\mathrm{p}$ are the
orbital periods.

If $m_2 \sim m_1$, the term with $e_\mathrm{ff}$ in the perturbing
function dominates. In the alternative case, when $m_2 \ll m_1$,
the dominating term is a secular one and the forced eccentricity
is

\begin{equation}
e_\mathrm{f} = \frac{5}{4} (1 - 2\mu)
\frac{a_\mathrm{b}}{a_\mathrm{p}} e_\mathrm{b} = \frac{5}{4} (1 -
2\mu) \left(\frac{T_\mathrm{b}}{T_\mathrm{p}}\right)^{2/3}
e_\mathrm{b} \label{eq_e_fP}
\end{equation}

\noindent \citep{MN04,DS15}.

As shown in \cite{DS15}, the secular time variation of the
planetary eccentricity, excited by the central binary, follows the
law $e_\mathrm{p} = 2 e_\mathrm{f} | \sin \omega t |$, where
$\omega$ is the variation frequency and $t$ is time. Therefore,
for the effective planetary eccentricity one can take the
time-averaged value $\langle e_\mathrm{p} \rangle = \frac{4}{\pi}
e_\mathrm{f}$ in the case of secular variations; and, analogously,
$\langle e_\mathrm{p} \rangle = \frac{4}{\pi} e_\mathrm{ff}$ in
the case of local time variations.

From formula~(\ref{eq_e_ff}), for the effective tidally excited
eccentricity of a CBP of a twin ($m_1 = m_2$) and circular
($e_\mathrm{b}=0$) binary one has

\begin{equation}
\langle e_\mathrm{p} \rangle = \frac{3}{4 \pi}
\left(\frac{a_\mathrm{b}}{a_\mathrm{p}}\right)^2 =
\frac{3}{4 \pi}
\left(\frac{T_\mathrm{b}}{T_\mathrm{p}}\right)^{4/3} .
\label{eq_e_ff_tw}
\end{equation}

Whichever the origin of the planetary eccentricity may be, the
tidal heating flux through the planetary surface is given by

\begin{equation}
h_\mathrm{p} = \frac{63}{16 \pi} {\cal G}^{3/2} M_\mathrm{s}^{5/2}
R_\mathrm{p}^3 Q_\mathrm{p}'^{-1} a_\mathrm{p}^{-15/2}
e_\mathrm{p}^2 \label{eq_h_tidal}
\end{equation}

\noindent \citep{JGB08,BJG09}, where ${\cal G}$ is the
gravitational constant, $M_\mathrm{s}$ is the stellar mass,
$a_\mathrm{p}$ and $e_\mathrm{p}$ are the semimajor axis and
eccentricity of the planetary orbit, $R_\mathrm{p}$ is the
planet's radius, and $Q_\mathrm{p}' = 3Q_\mathrm{p} /(2
k_\mathrm{p})$, where $Q_\mathrm{p}$ and $k_\mathrm{p}$ are the
planetary tidal dissipation parameter and Love number,
respectively. In our hierarchical model of a planet orbiting
around a close twin ($m_1 = m_2$) binary, one can set
$M_\mathrm{s}=m_1+m_2 =2m_1=2m_2$. Combining
Eq.~(\ref{eq_h_tidal}) with Eq.~(\ref{eq_e_ff_tw}), one has

\begin{equation}
h_\mathrm{p} = \frac{567}{256 \pi^3} \zeta {\cal G}^{3/2}
M_\mathrm{s}^{5/2} R_\mathrm{p}^3 Q_\mathrm{p}'^{-1}
a_\mathrm{b}^{4} a_\mathrm{p}^{-23/2} ,
\label{eq_h_tres1}
\end{equation}

\noindent where $\zeta$ is the unitless constant responsible for
the choice of a tidal model, as specified below.

The radius $a_\mathrm{cr}$ of the chaotic zone around a circular
twin binary can be estimated, using a numerical-fitting relation
derived in \cite{HW99}, as

\begin{equation}
a_\mathrm{cr} \approx 2.4 a_\mathrm{b} .
\label{eq_acr}
\end{equation}

\noindent Assuming (as justified in \citealt{S17AJ}; see also
discussion below) that the planet initially moves in a resonant
cell just at the edge of the central chaotic zone, one can express
$h_\mathrm{p}$ through the binary period $T_\mathrm{b}$:

\begin{equation}
h_\mathrm{p} = A_h {\cal G}^{-1} R_\mathrm{p}^3 Q_\mathrm{p}'^{-1}
T_\mathrm{b}^{-5} , \label{eq_h_Tb}
\end{equation}

\noindent where the unitless constant $A_h \approx \frac{567}{8}
2.4^{-23/2} \pi^2 \zeta \approx 0.0297 \zeta$.

\section{The escape process}
\label{sec_ep}

Both single-star and binary-star populations can lose their
planets by a number of dynamical and physical mechanisms: stellar
flybys, planet-planet scattering, supernova explosions, etc.; see
\cite{RF96,VT12,VEW14,K16}. No comparative planetary escape
statistics are available up to now, due to the complex and
multifaceted nature of the phenomenon. Concerning circumbinary
systems, \cite{SF16} found, based on massive numerical
simulations, that the planetary escape from such systems may
efficiently fill the Galaxy with rogue planets. According to
\cite{SF16,SKS16}, most CBP are ejected rather than destroyed via
planet-planet or planet-star collisions. Note that a major
difference of our tidal mechanism from the planet-planet
scattering is that it acts permanently, while the planet-planet
scattering is efficient only in young planetary systems.

Let us concentrate on circumbinary systems. A gravitating binary
with components of comparable masses possesses a circumbinary zone
of dynamical chaos if the mass ratio exceeds some threshold; this
zone of instability is formed by the overlap of the integer
mean-motion resonances (accumulating to the separatrix
corresponding to the parabolic motion) between the central binary
and the planet \citep{S15}. As follows from the orbital data on
the recently discovered CBP of main-sequence stars ({\it
Kepler}-16b, 34b, 35b, and others), most of them move in orbits
closely encircling the central chaotic zone \citep{D11,W12,PS13}.
In \cite{S15}, the extent of the chaotic zone around a system of
two gravitationally bound bodies was estimated analytically, based
on Chirikov's \citep{C79} resonance overlap criterion. The
binary's mass ratio, above which such a chaotic zone is
universally present, was also estimated as equal to $\approx
0.05$. Note that central cavities of analogous origin are observed
in protoplanetary disks, which contain planetesimals, dust, and
gas; the gas is present at the initial stages of the disk
evolution. The existence and properties of the central cavities in
gaseous circumbinary disks were first considered analytically in
\cite{AL94}. When CBP are formed, they migrate towards the central
stellar binary, and stall at the outer border of the chaotic zone
around the binary, because there is no more matter to cause the
migration \citep{PN07,M12,PLT12}.

After completing this relatively fast migration and being stalled
in a resonance cell just at the chaos border, is it possible for a
planet to complete the remaining part of the journey and to enter
the zone of chaos by any mechanism? As noted in \cite{S17AJ}, one
such mechanism may be provided by the slow stellar evolution of a
parent binary star. Various outcomes of the evolution are
possible; however, final supernova explosions may indeed free any
planetary material \citep{K16}.

As proposed in the Introduction, here we consider a different
mechanism of planetary escape---that of the tidal decay of
circumbinary orbits. Let us assess its efficiency. At a given
eccentricity $e_\mathrm{p}$, the size of a planetary orbit tidally
decays at the rate

\begin{equation}
\frac{1}{a_\mathrm{p}} \frac{d a_\mathrm{p}}{d t}  = -
\frac{63}{4} {\cal G}^{1/2} M_\mathrm{s}^{3/2} m_\mathrm{p}^{-1}
R_\mathrm{p}^5 Q_\mathrm{p}'^{-1} a_\mathrm{p}^{-13/2}
e_\mathrm{p}^2
\label{dap_tidal}
\end{equation}

\noindent (as can be derived from equations (1) and (2) in
\citealt{VBG14}).

Comparing Eqs.~(\ref{eq_h_tidal}) and (\ref{dap_tidal}), one has

\begin{equation}
\frac{1}{a_\mathrm{p}} \frac{d a_\mathrm{p}}{d t}  = - \frac{4 \pi
R_\mathrm{p}^2 a_\mathrm{p}} {{\cal G} M_\mathrm{s} m_\mathrm{p}}
h_\mathrm{p} . \label{dap_tidal_h}
\end{equation}

\noindent The formal characteristic timescale for the decay of a
planet's circumbinary orbit from its initial size $a_\mathrm{p}$
to zero (i.e., the variation in orbital radius $d a_\mathrm{p}
\sim a_\mathrm{p}$) is given by

\begin{equation}
\tau_0  = \frac{{\cal G} M_\mathrm{s} m_\mathrm{p}} {4 \pi
R_\mathrm{p}^2 a_\mathrm{p} h_\mathrm{p}} = \frac{{\cal G}
M_\mathrm{s} \rho_\mathrm{p} R_\mathrm{p}} {3 a_\mathrm{p}
h_\mathrm{p}} , \label{tau_0}
\end{equation}

\noindent where $\rho_\mathrm{p}$ is the density of the planet.

The time needed to traverse a half of an integer mean-motion
resonance cell at the border of the circumbinary chaotic zone is
an order of magnitude shorter. To prove this, let us consider two
neighboring integer and half-integer mean-motion resonances
$m$:$1$ and $(2m+1)$:$2$ between the binary and the planet; then,
the relative radial distance between the resonances is
$\frac{\Delta a}{a} \simeq \frac{2}{3(2m+1)}$ for $m \gg 1$.
Therefore, in the range of interest ($m=4, 5, 6$) one has
$\frac{\Delta a}{a} \simeq 0.07$--0.05. However, the chaotic
layers formed at the integer resonances near the chaos border are
broad, and this serves to diminish the distance needed to
traverse. (Recall that the measures of the chaotic and regular
components of phase space of the standard map at the critical
value of the stochasticity parameter are approximately equal; see
\citealt{S04}.) Therefore, we set $\xi \equiv \frac{\Delta a}{a}
\simeq 0.05$, noting that this estimate is approximate and may
even be an upper bound. This estimate is graphically confirmed,
e.g., by the stability diagram for {\it Kepler}-16 in
\cite[figure~3]{PS13}, where the planet {\it Kepler}-16b is
located just at the given distance from the nearest inner chaotic
layer.

Thus, for the upper bound of the timescale of traversing a half of
an integer mean-motion resonance cell at the border we adopt

\begin{equation}
\tau  = \xi \tau_0 ,
\label{tau_cell}
\end{equation}

\noindent where $\xi = 0.05$.

Again, we assume that the planet, after it has been formed and has
completed primordial migration, is stalled at the edge of the
circumbinary chaotic zone; therefore, $a_\mathrm{p} =
a_\mathrm{cr} \approx 2.4 a_\mathrm{b}$ (see
Eq.~(\ref{eq_acr})). Then, combining Eqs.~(\ref{eq_h_Tb}),
(\ref{tau_0}), and (\ref{tau_cell}), one can express $\tau$
through the binary period $T_\mathrm{b}$:

\begin{equation}
\tau  = B_\tau {\cal G}^{5/3} M_\mathrm{s}^{2/3} \rho_\mathrm{p}
R_\mathrm{p}^{-2} Q_\mathrm{p}' T_\mathrm{b}^{13/3} ,
\label{tau_cTb}
\end{equation}

\noindent where $M_\mathrm{s}=2m_1=2m_2$ ($m_1 = m_2$), and the
unitless constant $B_\tau = \frac{4^{1/3} \pi^{2/3} \xi}{3 \cdot
2.4 A_h} \simeq 15.9 \frac{\xi}{\zeta}$.

Finally in this Section, let us discuss in more detail the choice
of the initial relative radial location of a CBP at the chaos
border. Note, first of all, that the fitting function (a
polynomial fit), obtained in \cite{HW99}, is in fact a severely
smoothed version of the reality. The real border between chaotic
and regular domains in the phase space of motion is fractal; and
this manifests itself most clearly in charts of global dynamics,
e.g., in the ``pericentric distance -- eccentricity'' or
``semimajor axis -- eccentricity'' (initial values are implied)
diagrams; see \cite{PS13} and \cite{MW06} for the circumbinary and
the circumcomponent cases of such diagrams, respectively. In
particular, {\it Kepler}-16b appears to be literally ``immersed''
in the fractal border, formed by the Farey tree of the mean-motion
resonances between the central binary and the planet; see
\cite[figs.~2 and 3]{PS13}. Therefore, the real distance from the
close-to-border planet to the nearest instability layers can be
much smaller than the distance to the smoothed border, and thus
one may even assess the assumed 5\% limit as an upper bound.

Regarding the physical grounds for initially placing a CBP so
closely to the chaos border, it is important to note the
following.

Circumbinary gaseous protoplanetary disks are tidally truncated on
the inside \citep{AL94}; in modern numerical simulations the
radius of truncation $a_\mathrm{tr}$ of a disk around a binary
with $\mu \sim 1/2$ and $e_\mathrm{b} \sim 0$ is estimated to be
in the range from $\approx 2 a_\mathrm{b}$ \citep{SR15} to
$\approx 3 a_\mathrm{b}$ \citep{PPZ13}; i.e., it is taken to be
either somewhat smaller or somewhat greater than the radius of the
averaged (over the fractal structure, as discussed below) border
of the circumbinary chaotic zone $a_\mathrm{cr} \approx 2.4
a_\mathrm{b}$, given by Eq.~(\ref{eq_acr}) at $e_\mathrm{b} = 0$.
Therefore, at present it does not seem possible to derive expected
radii of the initial CBP pile-up with a high-enough precision,
based only on estimates of the radii of central cavities in
protoplanetary disks.

As initially hypothesized in \cite{S17AJ}, the escape process of
CBP may be eventually due to the orbital evolution of the host
stellar binary (due to stellar mass loss and mutual tides), and
not a CBP itself. In particular, at an early stage of a close
binary's evolution, due to the tidal transfer of the angular
momentum from the stellar rotation, the binary's orbit may widen
until the tidally-locked state (spin-orbit resonance 1:1) is
achieved. (At this moment, the size of the circumbinary chaotic
zone is maximum, as revealed in \citealt{FBG18}; see a note in
Section~``Conclusions.'') The pre-main-sequence (pre-MS) dwarf
binaries with periods less than 7--8~d are effectively
circularized by the stellar tidal interaction between companions,
and the most of the circularization takes place on a rather short
timescale, at the beginning of the Hayashi contraction stage
\citep{ZB89}. The Hayashi stage typically takes $\sim$1--10~Myr
\citep[eq.~(6)]{ZB89}, and the binary components reach the
tidally-locked state on a timescale much shorter than the
circularization timescale, as also revealed in \cite{ZB89}. On the
other hand, the lifetimes of circumbinary protoplanetary disks are
typically $\sim$1--10~Myr \citep{KIH12,PPZ13}. (The duration of
the disk era in the history of the Solar system was quite similar:
$\sim$10~Myr, see \citealt{MAC06}.) Therefore, one may expect that
the circumbinary chaotic zone is maximized already during the disk
era, and the formation and the primary (migrational) orbital
settlement of the CBP proceeds after (or marginally after) the
basic settlement of the parent binary's orbit. However, the
timescales may overlap, and the interplay of the processes is
possible. In our study we just assume that the parent binary's
orbit is already settled and is circular, and that its size is
constant.

\section{The lack of close binaries with CBP}

Now let us apply the developed theory and estimate numerically the
timescales of the effect in various possible planetary and tidal
heating models. Generally, close solar-type binaries with periods
less than 10 days are believed to be a product of some dissipative
evolution (most likely due to the Lidov--Kozai oscillations in
triples, with tidal friction), because they cannot be born in such
close pairs; see \cite{MK18} and references therein. On the other
hand, the existing correlation of companion masses and the
resemblance of the close-binary fractions for the pre-MS stars and
for the MS field stars testify that the evolution to the close
state is fast, taking $\lesssim 5$~Myr, most likely due to the
dissipation in primordial gas \citep{MK18}. Therefore, one can
consider the evolution of a circumbinary planetary system starting
from an early state of the binary that is already close.

More than half of all ($\sim 2000$) eclipsing binaries in the
first two quarters of the {\it Kepler} data have periods less than
7~d \citep[figure~8]{SPW11}, whereas all discovered {\it Kepler}
CBP (10 objects) have periods greater than 7~d. The dearth of CBP
of the binaries with $P \lesssim 5$~d was established as
statistically significant in \cite{AOB14,MT14}. This lack can be
interpreted as being due to the stellar Lidov--Kozai effect with
tidal friction, in the presence of an outer tertiary star
\citep{AOB14,MMF15,ML15,HPP16}. Conversely, the mechanism studied
here does not need a tertiary, although the process can be slower.

\begin{figure}[ht!]
\begin{center}
\includegraphics[width=6cm]{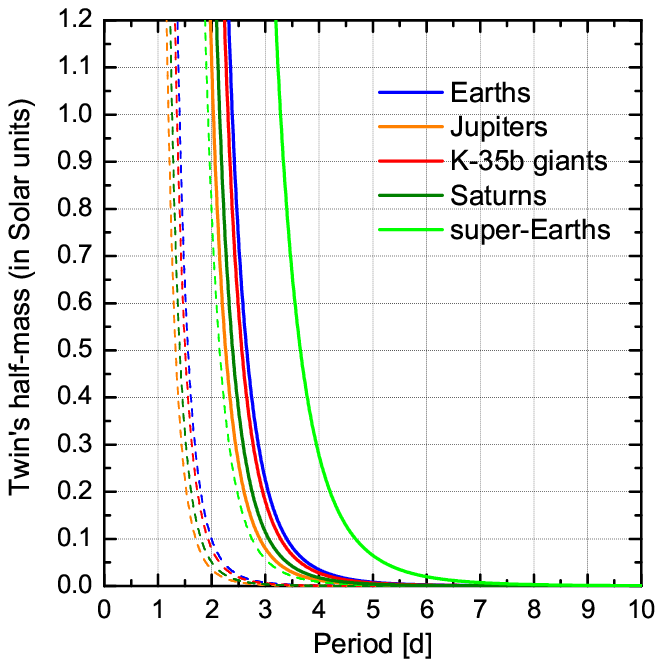}
\includegraphics[width=6cm]{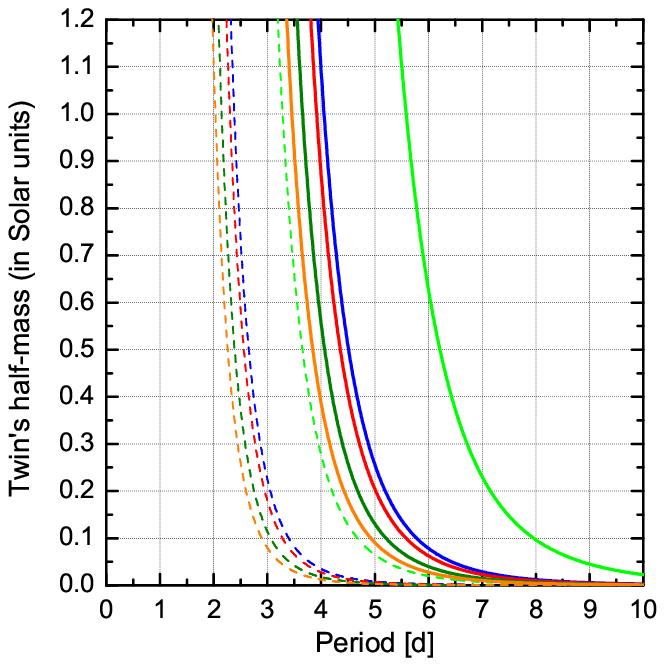}
\end{center}
\caption{Diagrams ``orbital period --- mass'' for twin stellar
binaries, with planetary escape isochrones superimposed. The mass
is counted for a companion, i.e., it is equal to a twin's
half-mass. The curves are the tidal escape isochrones
corresponding to escape times of 10$^{9}$~yr (left panel) and
10$^{10}$~yr (right panel) in various planetary models. The curves
are blue for Earth-like rocky planets, green for super-Earths,
olive for Saturn-like giants, orange for Jupiter-like giants, and
red for {\it Kepler}-35b-like giants. The dashed curves correspond
to $\zeta = 1$, the solid ones to $\zeta = 10$.} \label{fig1}
\end{figure}

To represent the effect graphically, we construct diagrams
``orbital period~--- mass'' for twin stellar binaries
(Fig.~\ref{fig1}). The mass is counted for a companion, i.e., it
is equal to a twin's half-mass. The depicted curves are the tidal
escape isochrones in several planetary models. The curves are
defined by Eq.~(\ref{tau_cTb}) with the escape time $\tau$ fixed
to 10$^{9}$ (1~billion) yr (in the left panel) and 10$^{10}$
(10~billion) yr (in the right panel). The curves for Earth-like
rocky planets are drawn in blue, those for super-Earths in green.
In the first case, we set $R_\mathrm{p} = R_\mathrm{Earth}$,
$\rho_\mathrm{p} = \rho_\mathrm{Earth}$, $Q_\mathrm{p}' = 100$, as
adopted in \cite{VBG14}. In the second case, the parameters are
the same except $R_\mathrm{p} = 2 R_\mathrm{Earth}$.

The curves in the diagrams are olive for Saturn-like giants,
orange for Jupiter-like giants, and red for {\it Kepler}-35b-like
giants. For Saturns, we set $R_\mathrm{p} = 9.07
R_\mathrm{Earth}$, $\rho_\mathrm{p} = 0.71$~g/cm$^3$; for
Jupiters, $R_\mathrm{p} = 11.0 R_\mathrm{Earth}$, $\rho_\mathrm{p}
= 1.33$~g/cm$^3$; and for {\it Kepler}-35b-like giants,
$R_\mathrm{p} = 0.728 R_\mathrm{Jupiter}$, $\rho_\mathrm{p} =
0.41$~g/cm$^3$ (as determined in \citealt{W12}). For all giants we
set $Q_\mathrm{p}' = 10^5$, as a typical value, according to
\cite{GS66}.

There exists a variety of models of planetary tidal heating.
According to \cite{RH18}, in the Andrade and Sundberg--Cooper
rheologies the tidal heating can be 10 times or even more greater
than in the traditional Maxwell model. To compare graphically the
role of the choice of tidal model, the curves for $\zeta=1$ in
both panels of Fig.~\ref{fig1} are dashed, and the curves for
$\zeta=10$ are drawn solid. We see that the solid isochrones are
somewhat to the right of the dashed ones, thus permitting more
planets to escape.

Any planet located on the left of a curve in the diagrams in
Fig.~\ref{fig1} would be liberated on a timescale that is less
than that corresponding to the curve. In total, we see that, on
timescales less than the current age of the Universe, the proposed
mechanism may explain, at least partially, the observed lack of
CBP of close-enough (with periods of several days) stellar
binaries. As follows from the diagrams, the mechanism is
especially effective for low-mass binaries and for super-Earths.

The increase in the tidal efficiency in the models of \cite{RH18}
can be assessed now only qualitatively, by the order of magnitude;
in fact, the efficiency can be {\it more} than ten times greater
than in the traditional models. Therefore, an increase of 35 times
that is needed to make the described process relevant for 10~Gyr
old solar-type binaries with periods of 5 days can be well within
reach of the new tidal models. In this framework, the fact that
many {\it Kepler} CBP are not orbiting right at the chaos border,
but some 10\% or further away, can be interpreted as a result of
the already-accomplished removal of closer planets, because the
tidal decay accelerates (being sharply dependent on the radial
distance) as the orbit shrinks.

Of course, the enhanced tidal efficiency is still debatable, but
what one can say for sure is that for the binaries with orbital
periods of two-three days or less the tidal decay can compete with
any other possible mechanism of the CBP removal, even for
solar-type binaries and even without any enhancement in the tidal
efficiency, as directly follows from the diagram in the right
panel of Fig.~1. What may be even more important, in the future,
is that it may remove more and more planets, as directly follows
from the same diagram.

\section{Post-CBP population of rogue planets}

The main-sequence binary stellar population in the Galaxy has a
rather broad distribution of periods \citep{DM91}, with a median
value at $\approx$180~yr. However, this distribution includes a
physically distinct stellar subpopulation, that of the so-called
``twin binaries,'' i.e. those with almost equal mass (with mass
ratios of the companions ranging from $\sim$0.8 to 1), which form
a statistical excess at short orbital periods
\citep{HMU03,Lu06,SO09}. For the twins, the median period is
$\sim$7~d, and the upper cut-off of the period distribution is at
$\approx$43~d \citep{Lu06,SO09}. Note that M dwarfs comprise the
majority of stars in the Galaxy, whereas binaries dominate in
number over single stars: M dwarfs comprise more than 70\% of the
Galactic stellar population, and more than 50\% of the stellar
population are in binaries \citep{DM91,BHC10}. Therefore, M-dwarf
twin binaries do not at all form an exotic class of objects. A
well known example of an M-dwarf twin binary located not far away
from the Solar system is given by EZ~Aqr A--C. Its potential CBP
were considered in \cite{PS16a} and are also discussed here
further on.

Let us estimate very roughly how many post-CBP could have already
been produced in the Galaxy, by means of the described mechanism.
Taking into account that the median period for twins is $\sim$7~d,
one may infer from inspection of the right panel of
Fig.~\ref{fig1} that the bulk of CBP could have already been (on a
timescale less than $10^{10}$~yr) liberated (thus becoming
post-CBP), only in the subclass of super-Earths orbiting low-mass
stars. As soon as the twin subpopulation represents $\sim$2--3\%
of all spectroscopic binaries \citep{Lu06,SO09}, and spectroscopic
binaries may represent $\gtrsim$5\% of all binaries
\citep[section~3.3.2]{DK13} one finds that the fraction of
potential post-CBP producers can be estimated as $\sim$0.1\% of
the total stellar population. Therefore, the total number of
already liberated CBP is at most $10^9$. This estimate at least
does not contradict a finding by \cite{MUS17} based on
observations of gravitational microlensing events that the
frequency of rogue (or wide-orbit) planets is less than 0.25
planets per main-sequence star. However, this latter upper limit
concerns Jupiter-mass objects; the statistics can be different for
smaller-mass rogue planets.

\section{Massive liberation of CBP}

\begin{figure}[ht!]
\begin{center}
\includegraphics[width=6cm]{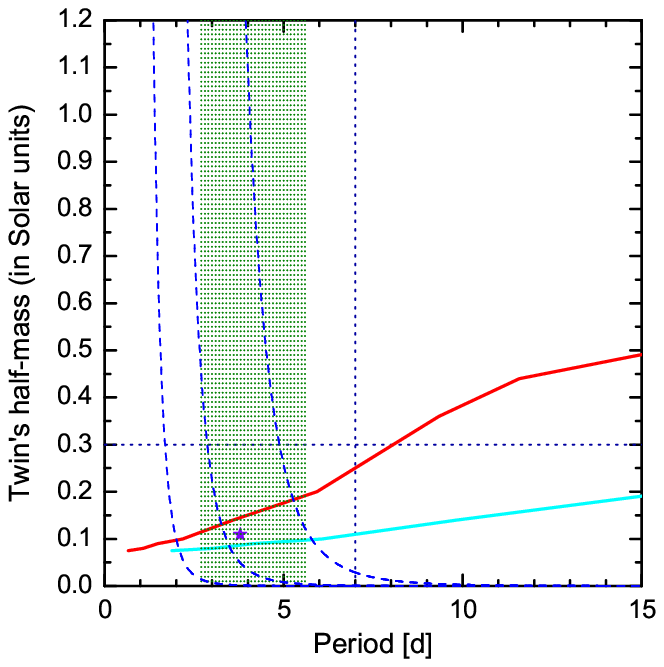}
\includegraphics[width=6cm]{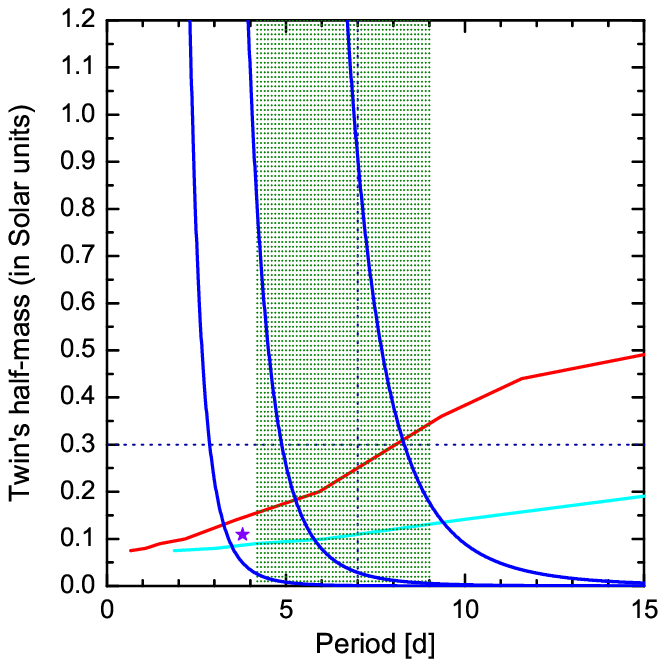}
\end{center}
\caption{Diagrams ``orbital period --- mass'' for twin stellar
binaries, with habitable zones (both insolation and tidal ones)
and planetary escape isochrones superimposed. The green-shaded
band is the tidal HZ for a CBP orbiting at the edge of the
circumbinary chaotic zone. Its vertical borders are specified by
formula~(\ref{eq_h_Tb}) with $h_\mathrm{p}=$ 0.04~W~m$^{-2}$ and
2~W~m$^{-2}$ (see text). The red and cyan curves are the ``hot''
and ``cold'' borders of the insolation HZ, drawn according to the
theory of \cite{KRK13}. The blue curves (both dashed and solid)
are the escape isochrones corresponding (from left to right in
each panel) to the escape times of 10$^{9}$, 10$^{10}$, and
10$^{11}$~yr. The horizontal and vertical dotted lines correspond
to the median values of masses and orbital periods for the
twin-binary subpopulation. The asterisk marks the location of
EZ~Aqr A--C. Left: the habitable zones and escape isochrones are
for a standard Earth-like rocky planet and $\zeta = 1$. Right: the
same except $\zeta = 10$.}
\label{fig2}
\end{figure}

Now let us consider longer timescales, which may permit massive
production of complex chemicals on Earth-like planets. In
Fig.~\ref{fig2}, diagrams ``orbital period --- mass'' are
constructed for a rocky Earth-like planet, with insolation and
tidal habitability zones (HZ) superimposed, to demonstrate how
they overlap with planetary escape isochrones. We adopt the
parameters for the standard Earth-like planet, as specified above.
To illustrate graphically the role of the choice of tidal model,
the diagram in the left panel of Fig.~\ref{fig2} is constructed
with $\zeta=1$, and that in the right panel with $\zeta=10$. It
should be emphasized that the parameters for rocky planets used to
construct both diagrams in Fig.~\ref{fig2} are standard and the
same. The difference is in the choice of the tidal dissipation
model.

According to heuristic estimates in \cite{BJG09}, heating rates of
less than 0.04~W~m$^{-2}$ and greater than 2~W~m$^{-2}$ imply
non-habitability. Thus, no biogenic chemicals are produced if a
planet does not fit this range. We assume that most of complex
planetogenic (not only biogenic) chemicals are produced on planets
that fit this range; heating rates somewhat higher than the given
upper bound can also be plausible, because they allow for active
tectonics, and thus a planet can be regarded as an active chemical
reactor.

The green-shaded vertical band in the diagrams in
Fig.~\ref{fig2} is the tidal HZ for a CBP orbiting at the edge of
the circumbinary chaotic zone. Its vertical borders are specified
by formula~(\ref{eq_h_Tb}) with $h_\mathrm{p}=$ 0.04~W~m$^{-2}$
and 2~W~m$^{-2}$, as referenced above. The red and cyan curves are
the ``hot'' and ``cold'' borders of the insolation HZ, drawn
according to the theory of \cite{KRK13} and the data on average
masses, temperatures, and luminosities of M-dwarf binaries from
\cite[Table~1]{KT09}.

The blue curves (both dashed and solid) are the tidal escape
isochrones corresponding (from left to right) to escape times of
10$^{9}$, 10$^{10}$, and 10$^{11}$~yr. The horizontal and vertical
dotted lines correspond to the median values of masses and orbital
periods for the twin-binary subpopulation. (According to
\citealt[figure~23]{BHC10}, the mass function of M dwarfs peaks at
classes M3V--M4V, i.e., at mass 0.20--0.36 in Solar units; see
also \citealt[Table~1]{KT09}.) We see that at $\zeta=10$ the line
of the median orbital period for the twin-binary subpopulation is
deep inside the tidal HZ.

The proposed mechanism provides the liberation timescales in any
range, depending on the closeness of the parent binary, and this
is an advantage. Indeed, other possible mechanisms act either in
young planetary systems (e.g., liberation due to planet-planet
scattering, see \citealt{RF96}) or in very old planetary systems
(e.g., liberation due to post-main-sequence loss of mass of parent
stars, see \citealt{VEW14}; or the liberation due to supernova
explosions of parent stars, see \citealt{K16}). Thus, the proposed
mechanism is able to liberate planets that are on the one hand
chemically evolved, and on the other hand not ruined by processes
of late stellar evolution.

In massive numerical simulations of the long-term dynamics of CBP
subject to slow inward migration, \cite{SF16} found that a
migrating planet entering a chaotic layer corresponding to an
integer mean-motion resonance (in the vicinity of the circumbinary
chaotic zone) with the central binary is liberated virtually
always without collision with any component of the parent binary.
Even if the initial conditions are chosen arbitrarily inside the
central chaotic zone, the collisionless outcome exceeds 80\%.
Therefore, whichever the mechanism of inward migration might be,
the process of liberation of a planet entering the circumbinary
chaotic zone is effectively non-ruinous for the planet.

\section{Discussion}
\label{sec_Discussion}

CBP can be mostly liberated and become rogue and then migrate
freely, whereas planetary systems of single stars are generally
expected to be stable (as the Solar system is, where only Mercury
can be ejected on a billion-year timescale, see
\citealt{L94,L08}). As described in \cite{S10}, the typical mode
of disruption of a hierarchical three-body system is a kind of
``L\'evy unfolding'' of the system in both time and space: at the
edge of the system's disruption, the escaping body exhibits L\'evy
flights in its orbital period and semimajor axis, and in the
course of this random process the orbital period and semimajor
axis become arbitrarily large until the separatrix between the
bound and unbound states of the motion is crossed and the body
escapes. Therefore, any escaping particle eventually moves close
to the ``parabolic separatrix.'' As the orbital velocity of a
particle in a parabolic trajectory about the barycenter is
proportional to $r^{-1/2}$ (where $r$ is the radial distance from
the barycenter), ``at infinity'' the escaping particle would have
an almost zero velocity with respect to the barycenter; a small
surplus is provided by the ``final kick'' from the parent binary,
which allows the planet to cross the separatrix. The diffusion of
chemically evolved planets inside the Galaxy may serve to
disseminate the planetogenic and biogenic chemicals even outside
the Galactic habitable zone. This annular concentric Galactic zone
has a rather restricted radial size, as revealed in \cite{LFG04}.

Why should one care about the chemically evolved status of the
ejected planets? It is plausible to assume that planetary
chemicals are indeed necessary in some way for the long-lasting
galactic chemical evolution, as, analogously, stellar-born metals
are necessary. Therefore, (1)~the ejected planets must be
chemically evolved in some sense (enriched by complex chemicals),
(2)~the ejection process must keep the produced matter safe. Both
these conditions are obviously satisfied in the proposed scenario,
because the process is slow enough and mostly collisionless. As we
know from the sole example of Earth, the evolution to, e.g., the
superhabitability stage (as defined in \citealt{HA14}), associated
with the maximum in biomass production, may take some billion
years. Therefore, it may need gigayears for a planet to become
chemically (or biochemically) evolved.

Similar massive processes in other galaxies may depend on their
type. In 1969, an excess of flux (the so-called {\it UV~upturn})
was discovered in the far-ultraviolet spectrum of elliptical
galaxies, in the course of pioneering observations of the bulge of
M31 from the {\it Stargazer} Orbiting Astronomical Observatory~2
\citep{Co69}. As later on suggested and argued in \cite{HPL07} and
\cite{HPL09}, the UV~upturn is caused by an old population of hot
helium-burning binary stars that lost their hydrogen-rich
envelopes due to binary interactions. This reveals the presence of
a large number of short-period binaries (and presumably
circumbinary planetary systems) in elliptical galaxies. Taking
into account that elliptical galaxies contain many more stars (by
one or two orders of magnitude) than spiral ones, one comes to the
conclusion that the described mechanism of production of
chemically evolved rogue planets may occur on far greater scales
in typical elliptic galaxies than in our Galaxy.

Habitability properties of CBP are of especial interest. As argued
in \cite{S17AJ}, striking analogies exist between the habitability
conditions on CBP and on the Earth. In its habitability
properties, the Earth seems to mimic a typical CBP in certain
classes of stellar binaries: according to \cite{S17AJ}, a system
consisting of a planet orbiting at the edge of the circumbinary
chaotic zone around a typical M-dwarf twin binary is a photo-tidal
equivalent of the Sun--Earth--Moon system. In this respect, of
particular interest in the diagrams in Fig.~\ref{fig2} is the
quadrangle formed by the boundaries of the two intersecting bands
corresponding to the tidal and insolation habitable zones. CBP of
any binaries in this quadrangle are favorable for the production
of biogenic chemicals. Tantalizingly, one can see that (1)~this
area tends to overlap with the maximum of the expected
distribution of the twin-binary subpopulation (this maximum is at
the intersection of the horizontal and vertical dotted lines,
corresponding to the median values of masses and orbital periods
for twin binaries); (2)~it is located between the escape
isochrones corresponding to escape times of 10$^{9}$ and
10$^{11}$~yr, which are long enough for bio-evolution but shorter
than stellar lifetimes. Indeed, the main-sequence lifetime of M
dwarfs is up to $10^{13}$ (ten trillion)~yr \citep{LBA97}.

The asterisk in the diagrams marks the location of the M-dwarf
binary EZ~Aqr A--C, as specified by the values of orbital period
and mass taken from \cite{DFU99}. According to \cite{PS16a}, a CBP
orbiting at the edge of the circumbinary chaotic zone of this
binary (which is, by the way, a close neighbor of the Sun) can
belong to the insolation HZ. Here we see that it can also belong
to the tidal HZ; moreover, the timescale of escape of such a
planet can be as short as $10^{9}$~yr.

What are the general observable signatures of the considered
effect? They can be basically twofold, comprising its
manifestations in statistical and physical properties of
(1)~circumbinary planetary systems and (2)~rogue planets. In the
first case, the effect can provide the lack of observed CBP
belonging to close-enough binaries, as described above. In the
second case, one should underline that the ejected planets
interact with field stars (for a review about similar stellar
interactions, see \citealt[section~3.1]{YuT03}), and specifically
with binaries. They can even be emporarily captured by binaries,
analogously to the process of capture of dark matter particles by
binaries, considered in \cite{RLS15}. Concerning observations of
individual rogue planets, post-CBP may indeed have specific
physical properties that are potentially observable \citep{SF16}.

An inspection of Eq.~(\ref{tau_cTb}) and Fig.~(\ref{fig1}) makes
it clear that the shortest liberation timescales are achieved on
the one hand for large rocky planets (super-Earths), and on the
other hand for CBP in systems of low-mass binaries (the lower a
binary's mass, the shorter the timescale). This allows one to
conclude that observational signatures of the effect may comprise:
(1)~a prevalence of super-Earths in the whole population of rogue
planets (of course, if this mechanism were the only source of
rogue planets); (2)~a mass-dependent paucity of CBP in systems of
low-mass binaries: the lower the stellar mass, the greater the
paucity. Note that there already exists observational evidence for
a moderate abundance of rogue super-Earths; see \cite{MUS17}.

An important distinctive property of the proposed mechanism is
that it is not expected to fill interstellar space with small
bodies, such as interstellar planetesimals. Indeed, according to
Eq.~(\ref{tau_cTb}), the escape timescale is proportional to a
circumbinary particle's size to the power minus two, and this
cannot be balanced by any realistic correlation with
$\rho_\mathrm{p}$ or $Q_\mathrm{p}'$. On the other hand, due to
variations of the planetary Hill radius with the distance from
either of the two parent stars, an escaping giant planet can in
parallel release dozens of its irregular satellites. Therefore,
production of objects such as 1I/'Oumuamua is not excluded.

It is worth noting that the tidal mechanism for escape of
circumbinary material, described in this article, can be also
relevant to astrophysical systems other than planetary ones. For
example, relevant orbital configurations of circumbinary
``satellites'' may occur around binary asteroids or TNOs, around
rotating contact binaries (such as cometary nuclei, many
asteroids, and TNOs), and around binary black holes. Analysis of
the efficiency of the mechanism in such different settings is
beyond the scope of this paper, but it can be worth studying.

\section{Conclusions}

We have shown that circumbinary planetary systems are subject to
universal tidal decay (shrinkage of orbits), caused by the forced
orbital eccentricity inherent to them. CBP are ejected from parent
systems when they enter the circumbinary chaotic zone.

On shorter timescales (less than the current age of the Universe),
the proposed effect may explain, at least partially, the observed
lack of CBP of close-enough (with periods $< 5$~days) stellar
binaries. On longer timescales (greater than the age of the
Universe but well within stellar lifetimes), it may provide
massive liberation of chemically evolved CBP.

It should be underlined that the phenomenon of tidal decay of
circumbinary systems is universal. Here we have considered
planetary systems, but in fact it is present in any circumbinary
system of gravitating bodies.

While this paper was in the reviewing process, a preprint was
posted and an article published \citep{FBG18}, where the lack of
close isolated binaries with CBP was also explained by the
destabilization of CBP orbits due to their entering the
circumbinary chaotic zone. However, the mechanism of
the entry is different from that considered above.
In our scenario, the planetary orbit slowly
shrinks, while the chaotic zone stays constant in size. In the
tidal scenario of \cite{FBG18}, in contrast, the chaotic zone
swells (as the binary's orbit widens, due to the tidal transfer of
angular momentum from the stellar rotation), while the planetary
orbit stays constant in size. It should be noted that the tidal
scenario of \cite{FBG18} is relevant to an early
(pre-main-sequence) stage of evolution of the host star (see
Section~\ref{sec_ep}), whereas our scenario acts on much longer
timescales, and is thus capable of providing the escape of
chemically evolved planets.

\section*{Acknowledgements}

The author is grateful to the referee for
useful remarks and comments. This work was supported in part by
the Russian Foundation for Basic Research (project No.
17-02-00028) and by Programme~2 of Fundamental Research of the
Russian Academy of Sciences. In Section~7, the work was partially
supported by the Russian Scientific Foundation (project No.
16-12-00071).

\newpage

\begin{thebibliography}{}

\bibitem[Andrade-Ines \& Robutel(2017)]{AIR17}
Andrade-Ines, E., \& Robutel, Ph. 2018, Celest. Mech. Dyn. Astron.,
130, 6

\bibitem[Armstrong et al.(2014)]{AOB14}
Armstrong, D. J., Osborn, H., Brown, D., et al. 2014, MNRAS, 444,
1873

\bibitem[Artymowicz \& Lubow(1994)]{AL94}
Artymowicz, P., \& Lubow, S. H. 1994, ApJ, 421, 651

\bibitem[Barnes et al.(2009)]{BJG09}
Barnes, R., Jackson, B., Greenberg, R., \& Raymond, S. N. 2009,
ApJ, 700, L30

\bibitem[Bochanski et al.(2010)]{BHC10}
Bochanski, J. J., Hawley, S. L., Covey, K. R., et al. 2010, AJ,
139, 2679

\bibitem[Chirikov(1979)]{C79}
Chirikov, B. V. 1979, Phys. Rep., 52, 263

\bibitem[Code(1969)]{Co69}
Code, A. D. 1969, Publ. Astron. Soc. Pacific, 81, 475

\bibitem[Delfosse et al.(1999)]{DFU99}
Delfosse, X., Forveille, T., Udry, S., et al. 1999, A\&A, 350, L39

\bibitem[Demidova \& Shevchenko(2015)]{DS15}
Demidova, T. V., \& Shevchenko, I. I. 2015, ApJ, 805, 38

\bibitem[Doyle et al.(2011)]{D11}
Doyle, L., Carter, J. A., Fabrycky, D. C., et al. 2011, Science,
333, 1602

\bibitem[Duch\^{e}ne \& Kraus(2013)]{DK13}
Duch\^{e}ne, G., \& Kraus, A. 2013, ARA\&A, 51, 269

\bibitem[Duquennoy \& Mayor(1991)]{DM91}
Duquennoy, A., \& Mayor, M. 1991, A\&A, 248, 485

\bibitem[Fleming et al.(2018)]{FBG18}
Fleming, D. P., Barnes, R., Graham, D. E., Luger, R., \& Quinn, T. R. 2018, ApJ, 858, 86

\bibitem[Goldreich \& Soter(1966)]{GS66}
Goldreich, P., \& Soter, S. 1966, Icarus, 5, 375

\bibitem[Halbwachs et al.(2003)]{HMU03}
Halbwachs, J. L., Mayor, M., Udry, S., \& Arenou, F. 2003, A\&A,
397, 159

\bibitem[Hamers, Perets, \& Portegies Zwart(2016)]{HPP16}
Hamers, A. S., Perets, H. B., \& Portegies Zwart, S. F. 2016,
MNRAS, 455, 3180

\bibitem[Han, Podsiadlowski, \& Lynas-Gray(2007)]{HPL07}
Han, Z., Podsiadlowski, Ph., \& Lynas-Gray, A. E. 2007, MNRAS,
380, 1098

\bibitem[Han, Podsiadlowski, \& Lynas-Gray(2009)]{HPL09}
Han, Z., Podsiadlowski, Ph., \& Lynas-Gray, A. E. 2009, in
Ch\'avez Dagostino, M., et al. (eds.), {\it New Quests in Stellar
Astrophysics II.} (Springer.) P.~59

\bibitem[Heller \& Armstrong(2014)]{HA14}
Heller, R., \& Armstrong, J. 2014, Astrobiology, 14, 50

\bibitem[Holman \& Wiegert(1999)]{HW99}
Holman, M. J., \& Wiegert, P. A. 1999, AJ, 117, 621

\bibitem[Jackson, Greenberg, \& Barnes(2008)]{JGB08}
Jackson, B., Greenberg, R., \& Barnes, R. 2008, ApJ, 681, 1631

\bibitem[Kaltenegger \& Traub(2009)]{KT09}
Kaltenegger, L., \& Traub, W. A. 2009, ApJ, 698, 519

\bibitem[Kopparapu et al.(2013)]{KRK13}
Kopparapu, R. K., Ramirez, R., Kasting, J. F., et al. 2013, ApJ,
765, 131

\bibitem[Kostov et al.(2016)]{K16}
Kostov, V. B., Moore, K., Tamayo, D., Jayawardhana, R., \&
Rinehart, S. A. 2016, ApJ, 832, 183

\bibitem[Kraus et al.(2012)]{KIH12}
Kraus, A. L., Ireland, M. J., Hillenbrand, L. A., \& Martinache,
F. 2012, ApJ, 745, 19

\bibitem[Laskar(1994)]{L94}
Laskar, J. 1994, A\&A, 287, L9

\bibitem[Laskar(2008)]{L08}
Laskar, J. 2008, Icarus, 196, 1

\bibitem[Laughlin, Bodenheimer, \& Adams(1997)]{LBA97}
Laughlin, G., Bodenheimer, P., \& Adams, F. C. 1997, ApJ, 482, 420

\bibitem[Lineweaver, Fenner, \& Gibson(2004)]{LFG04}
Lineweaver, C. H., Fenner, Y., \& Gibson, B. K. 2004, Science,
303, 59

\bibitem[Lucy(2006)]{Lu06}
Lucy, L. B. 2006, A\&A, 457, 629

\bibitem[Martin, Mazeh, \& Fabrycky(2015)]{MMF15}
Martin, D. V., Mazeh, T., \& Fabrycky, D. C. 2015, MNRAS, 453,
3554

\bibitem[Martin \& Triaud(2014)]{MT14}
Martin, D. V., \& Triaud A. H. M. J. 2014, A\&A, 570, A91

\bibitem[Meschiari(2012)]{M12}
Meschiari, S. 2012, ApJ, 752, 71

\bibitem[Moe \& Kratter(2018)]{MK18}
Moe, M., \& Kratter, K. M. 2018, ApJ, 854, 44

\bibitem[Montmerle et al.(2006)]{MAC06}
Montmerle, Th., Augereau, J.-Ch., Chaussidon, M., et al. 2006,
Earth, Moon, and Planets, 98, 39

\bibitem[Moriwaki \& Nakagawa(2004)]{MN04}
Moriwaki, K., \& Nakagawa, Y. 2004, ApJ, 609, 1065

\bibitem[Mr\'oz et al.(2017)]{MUS17}
Mr\'oz, P., Udalski, A., Skowron, J., et al. 2017, Nature, 548,
183

\bibitem[Mudryk \& Wu(2006)]{MW06}
Mudryk, L. R., \& Wu, Y. 2006, ApJ, 639, 423

\bibitem[Mu\~{n}oz \& Lai(2015)]{ML15}
Mu\~{n}oz, D. J., \& Lai, D. 2015, Proc. Nat. Acad. Sci., 112,
9264

\bibitem[Paardekooper et al.(2012)]{PLT12}
Paardekooper, S.-J., Leinhardt, Z. M., Th\'ebault, T., \&
Baruteau, C. 2012, ApJ, 754, L16

\bibitem[Pelupessy \& Portegies Zwart(2013)]{PPZ13}
Pelupessy, F. I., \& Portegies Zwart, S. 2013, MNRAS, 429, 895

\bibitem[Pierens \& Nelson(2007)]{PN07}
Pierens, A., \& Nelson, R. P. 2007, A\&A, 472, 993

\bibitem[Popova \& Shevchenko(2013)]{PS13}
Popova, E. A., \& Shevchenko, I. I. 2013, ApJ, 769, 152

\bibitem[Popova \& Shevchenko(2016)]{PS16a}
Popova, E. A., \& Shevchenko, I. I. 2016, Astron. Lett., 42, 260

\bibitem[Rasio \& Ford(1996)]{RF96}
Rasio, F. A., \& Ford, E. B. 1996, Science, 274, 954

\bibitem[Renaud \& Henning(2018)]{RH18}
Renaud, J. P., \& Henning, W. G. 2018, ApJ, 857, 98

\bibitem[Rollin, Lages, \& Shepelyansky(2014)]{RLS15}
Rollin, G., Lages, J., \& Shepelyansky, D. L. 2015, A\&A, 576, A40

\bibitem[Shevchenko(2004)]{S04}
Shevchenko, I. I. 2004, JETP Letters, 79, 523

\bibitem[Shevchenko(2010)]{S10}
Shevchenko, I. I. 2010, Phys. Rev. E, 81, 066216

\bibitem[Shevchenko(2015)]{S15}
Shevchenko, I. I. 2015, ApJ, 799, 8

\bibitem[Shevchenko(2017)]{S17AJ}
Shevchenko, I. I. 2017, AJ, 153, 273

\bibitem[Silsbee \& Rafikov(2015)]{SR15}
Silsbee, K., \& Rafikov, R. R. 2015, ApJ, 808, 58

\bibitem[Simon \& Obbie(2009)]{SO09}
Simon, M., \& Obbie, R. C. 2009, AJ, 137, 3442

\bibitem[Slawson et al.(2011)]{SPW11}
Slawson, R. W., Pr\v{s}a, A., Welsh, W. F., et al. 2011, AJ, 142,
160

\bibitem[Smullen, Kratter, \& Shannon(2016)]{SKS16}
Smullen, R. A., Kratter, K. M., \& Shannon, A. 2016, MNRAS, 461,
1288

\bibitem[Sutherland \& Fabrycky(2016)]{SF16}
Sutherland, A. P., \& Fabrycky, D. C. 2016, ApJ, 818, 6

\bibitem[Van Laerhoven, Barnes, \& Greenberg(2014)]{VBG14}
Van Laerhoven, C., Barnes, R., \& Greenberg, R. 2014, MNRAS, 441,
1888

\bibitem[Veras \& Tout(2012)]{VT12}
Veras, D., \& Tout, C. A. 2012, MNRAS, 422, 1648

\bibitem[Veras et al.(2014)]{VEW14}
Veras, D., Evans, N. W., Wyatt, M. C., \& Tout, C. A. 2014, MNRAS,
437, 1127

\bibitem[Welsh et al.(2012)]{W12}
Welsh, W. F., Orosz, J. A., Carter, J. A., et al. 2012, Nature,
481, 475

\bibitem[Yu \& Tremaine(2003)]{YuT03}
Yu, Q., \& Tremaine, S. 2003, ApJ, 599, 1129

\bibitem[Zahn \& Bouchet(1989)]{ZB89}
Zahn, J.-P., \& Bouchet, L. 1989, A\&A, 223, 112

\end{thebibliography}
\end{document}